\documentclass[prl,twocolumn,showpacs]{revtex4}

\usepackage{graphicx}
\usepackage{amsmath}

\begin{document}

\title{Excitations of a Superfluid in a 3D Optical Lattice}
\author{Christian Schori$^{\dag }$, Thilo St\"{o}ferle, Henning Moritz,
Michael K\"{o}hl and Tilman Esslinger}
\affiliation{Institute of Quantum Electronics, ETH Z\"{u}rich H\"{o}nggerberg, CH--8093 Z%
\"{u}rich, Switzerland}
\date{\today}

\begin{abstract}
We prepare a Bose-Einstein condensed gas in a three-dimensional optical
lattice and study the excitation spectrum of the superfluid phase for
different interaction strengths. We probe the response of the system by
modulating the depth of the optical lattice along one axis. The interactions
can be controlled independently by varying the tunnel coupling along the
other two lattice axes. In the weakly interacting regime we observe a small
susceptibility of the superfluid to excitations, while for stronger
interactions an unexpected resonance appears in the excitation spectrum. In
addition we measure the coherent fraction of the atomic gas, which
determines the depletion of the condensate.
\end{abstract}

\maketitle

A variety of intriguing macroscopic quantum phenomena have become accessible
to experiments with quantum gases by loading them into the periodic
potential of an optical lattice. Initial experiments were carried out in the
weakly interacting regime which led to the observation of Josephson type
oscillations \cite{Anderson1998,Cataliotti2001}, analogous to those seen in
superconductors or superfluids. The strongly interacting regime could be
reached in experiments with extremely deep lattice potentials leading to
number squeezing \cite{Orzel2001,Hadzibabic2004}, and with three-dimensional
optical lattices leading to the observation of a quantum phase transition
from a superfluid to a Mott insulator \cite{Greiner2002a,Stoeferle2004}. In
the latter experiments the repulsive interaction energy $\ U$ between two
atoms within a minimum of the optical lattice potential is much larger than
the kinetic energy associated with the tunnel coupling $J$ between adjacent
minima \cite{Fisher1989,Freericks1994,Kuehner1998,Jaksch1998}.

In this paper we explore the excitation spectrum of a Bose-Einstein
condensed gas in an three-dimensional optical lattice where we can
continuously tune the experimental regime from weak ($U\ll J$) to
increasingly stronger ($U\approx J$) interactions. Modulation of the depth
of the optical lattice along one axis is used to excite the system while the
effect of interactions is controlled by increasing the potential depth along
the other two axes.

The strength of the excitation is measured by the fraction of atoms which
are exited out of the condensate. The weakly interacting Bose gas shows a
very low susceptibility to the modulation. This observation is in agreement
with a theoretical prediction derived with stationary Bogoliubov theory \cite%
{Menotti2003}. When we continuously increase $U/J$ we observe a gradual
appearance of an unexpected resonance in the excitation spectrum. For each
value of $U/J$ we measure the coherent fraction of the atomic gas to
determine the depletion of the condensate.

Our experimental setup is described in Ref. \cite{Moritz2003}. In brief, we
produce almost pure condensates of typically $1.5\times 10^{5}$ $^{87}$Rb
atoms in the hyperfine groundstate $|F=2,m_{F}=2\rangle $ which is confined
by a magnetic trap with trapping frequencies $\omega _{x}=2\pi \times 18\,%
\text{Hz}$, $\omega _{y}=2\pi \times 20\,\text{Hz}$, and $\omega _{z}=2\pi
\times 22\,\text{Hz}$. Three retro-reflected laserbeams (wavelength $\lambda
=826$ nm) are focused onto the condensate to form the optical lattice. At
the position of the condensate the gaussian shaped beams have $1/e^{2}$%
-radii of $120$\thinspace $\mu \text{m}$ ($x$ and $y$ axes) and $105$%
\thinspace $\mu \text{m}$ ($z$). The Thomas-Fermi radius of the magnetically
trapped condensate is $R_{TF}=13$ $\mu $m and the number of occupied sites
along the probe axis can be estimated to be $L=4R_{TF}/\lambda \simeq 60$.
For the maximal depth of the optical \ lattice potential the confinement of
the atomic cloud is dominated by the gaussian intensity profile of the
lattice beams which leads to a nearly isotropic harmonic potential with a
trapping frequency of $2\pi \times 47$ Hz.

\begin{figure}[tp]
\includegraphics[width=0.85\columnwidth,clip=true]{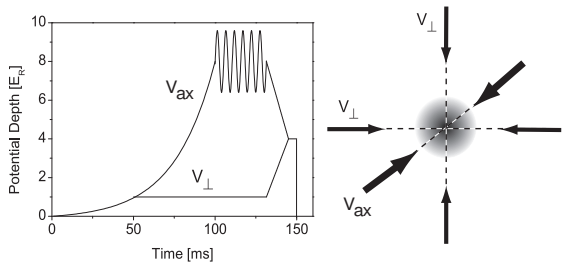}
\caption{An illustration of the experimental sequence used to measure the
excitation spectrum of the Bose gas in the optical lattice is shown on the
left. The optical lattice geometry is depicted on the right. The excitation
spectra are recorded as a function of the transferred energy $h{\protect\nu %
_{mod}}$ and repeated for different values of the transverse lattice depth $%
V_{\perp }$, see text.}
\label{fig1}
\end{figure}

To load the condensate into the ground state of the optical lattice, the
intensities of the lasers are slowly increased to their final values using
an exponential ramp with a time constant of $25\,\text{ms}$ and a duration
of $100\,\text{ms}$. The resulting optical potential depths $V_{y}\equiv
V_{ax}$ and $V_{x,z}\equiv V_{\perp }$ are proportional to the laser
intensities and are conveniently expressed in terms of the recoil energy $%
E_{R}=\frac{\hbar ^{2}k^{2}}{2m}$ with $k=\frac{2\pi }{\lambda }$ and the
atomic mass $m$. The specific loading sequence is shown in figure \ref{fig1}%
. The axial potential is ramped to $V_{ax}=8\,E_{R}$ and simultanously the
potential along the orthogonal axes is ramped to different values in the
interval $V_{\perp }=0..8\,E_{R}$. The ratio $U/J$ is set by the final value
of $\ V_{\perp }$. Here $J$ is the combined tunnel coupling between a single
site and all nearest neighbor sites, i.e. $J=2J_{x}+2J_{y}+2J_{z}$ where $%
J_{l}$ is the tunnel coupling between two adjacent sites along the axis $l$.

After loading the condensate into the optical lattice we modulate the
amplitude of the standing wave along the probe axis at a frequency ${\nu
_{mod}}$ (see figure \ref{fig1}). This modulation adds two sidebands at
frequency $\pm {\nu _{mod}}$ to the carrier frequency of the standing wave
field. Accordingly, two-photon Raman transitions can be induced which
transfer energy $h{\nu _{mod}}$, in analogy with Bragg spectroscopy on
magnetically trapped condensates \cite{Stenger1999}. For an extended
periodic system the quasi-momentum transfer $q$ of the modulation would be
zero. Due to the finite size of the sample the transferred quasi-momentum
lies in the interval $\delta q\approx \pm 2\hbar k/L$ around zero.

Following the excitation we ramp down the lattice potentials linearly in $15$
ms to $V_{ax}=V_{\bot }=4\,E_{R}$. The system is kept at this lattice depth
for $5$ ms to allow for re-thermalization \cite{thermalization}. Then all
optical and magnetic potentials are suddenly switched off. The resulting
matter wave interference pattern is detected by absorption imaging after $%
25\,\text{ms}$ of ballistic expansion. The duration $t_{mod}=30\,\text{ms}$
and amplitude $A_{mod}=0.2V_{ax}$ of the modulation are chosen such that we
always observe a finite condensate fraction. We have also verified that all
atoms remain in the lowest Bloch band by adiabatically switching off the
lattice potentials \cite{Greiner2001b} after the modulation.

\begin{figure}[tp]
\includegraphics[width=0.85\columnwidth,clip=true]{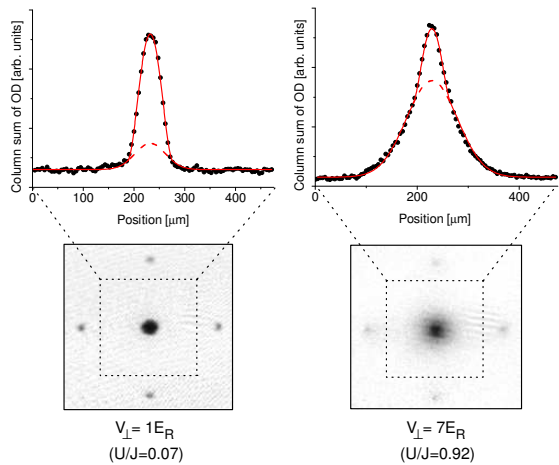}
\caption{Absorption images of the Bose gas taken after 25 ms of ballistic
expansion. Prior to the expansion the optical lattice depth along the probe
axis is modulated for a period of 30 ms at the frequency $\protect\nu %
_{mod}=3$ kHz . The data for the weakly interacting regime is displayed on
the left and the data for stronger interactions are shown on the right. The
total field of view of the images is $720$ $\protect\mu $m $\times $ $720$ $%
\protect\mu $m, solid box. The optical density of the central momentum peak
is fitted with a bimodal distribution from which the strength of excitation
is obtained, see text. The fit (solid line) and the column sum of the
optical density (circles) are shown in the upper row. The dashed line is the
gaussian part of the bimodal fit. The ratios $U/J$ are calculated
numerically \protect\cite{Jaksch1998}.}
\label{fig2}
\end{figure}

In figure \ref{fig2} we show absorption images taken after modulating the
optical lattice amplitude at a frequency ${\nu _{mod}}=3$ kHz. During the
modulation energy is transferred to the atomic gas which results in a
broadening of the central momentum peak. We analyze the central momentum
peak using a two dimensional bimodal distribution consisting of a gaussian
function for the thermal fraction and an inverted parabolic function for the
condensate component. We use the thermal fraction $1-N_{c}/N$ obtained from
the image analysis as a measure for the excitation strength, with $N_{c}$
and $N$ being the condensate and total number of atoms in the central
momentum peak. For the thermal fraction we find a minimum of around 0.2 even
when we reduce the amplitude of the modulation to zero. When we
adiabatically increase and subsequently reduce the lattice potential we do
not observe any significant heating of the condensate. However, during
expansion from the optical lattice atom-atom scattering may reduce the
condensate fraction. Also, when the condensate fraction is high the optical
density of the central momentum peak saturates and we underestimate the
number of atoms in the condensate. Finally, when the thermal fraction is
small the atomic distribution is no longer accurately fitted by the bimodal
distribution.

\begin{figure}[tp]
\includegraphics[width=1\columnwidth,clip=true]{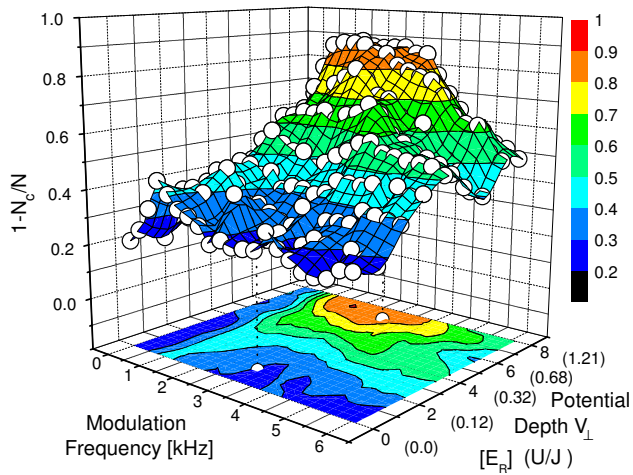}
\caption{Excitation strength of the Bose gas as a function of the transverse
lattice depth $V_{\perp }$ and the excitation frequency ${\protect\nu _{mod}}
$. The interaction ratios $U/J$ are given in brackets. The surface and
contour plot is generated from a Renka-Cline interpolation of approximately
160 data points (o). Data points from figure \protect\ref{fig2} are also
shown as projections onto the contour plane. The maximal observed scattering
among repeated measurements of the excitation strength is $\pm 0.02$.}
\label{fig3}
\end{figure}

Figure \ref{fig2} shows that the thermal fraction remains close to its
minimum when the modulation is performed in the weakly interacting regime ($%
V_{\perp }=1\,E_{R}$), i.e. the system is not susceptible to the excitation.
In the more strongly interacting regime ($V_{\perp }=7\,E_{R}$) the thermal
fraction approaches unity. Here the superfluid shows an increased
susceptibility to the modulation. We have measured a full series of spectra
to characterize the Bose gas in the transition region between weak and
increasingly stronger interactions. Each spectrum is recorded by scanning
the frequency ${\nu _{mod}}$ of the modulation for a fixed value of $%
V_{\perp }$. These data are displayed in figure \ref{fig3} where the
excitation strength is plotted as a function of the transverse lattice
potential $V_{\perp }$ and the frequency ${\nu _{mod}}$.

We first discuss our results for the weakly interacting Bose gas ($V_{\perp
}=0\,E_{R}$, and $U/J\ll 1$). In this regime the observed excitation
strength remains close to the lower detection limit of 0.2 for all
excitation energies. This observation can be related to the behaviour of the
static structure factor $S(q)$ derived from stationary Bogoliubov theory.
The structure factor describes the total strength of all possible
excitations with momentum transfer $q$. For $q$ close to zero the excitation
energy to the second Bogoliubov band is above $4E_{R}$ where $E_{R}/h=3.4$
kHz. Since the highest modulation frequency is $6$ kHz we can safely neglect
excitations beyond the lowest band. For long wavelength excitations ($q$
close to zero) the structure factor behaves like $S(q)\sim \varepsilon
(q)/\hbar \omega (q)$, where $\varepsilon (q)$ and $\omega (q)$ are the
lowest Bloch- and Bogoliubov bands in the optical lattice \cite{Menotti2003}%
. Due to the quadratic (linear) dispersion relation of the Bloch
(Bogoliubov) band near zero quasimomenta we expect a strong suppression of
the measured excitation strength for all excitation energies which is in
reasonable agreement with our experimental observation.

The excitation strength in figure \ref{fig3} shows the gradual appearance of
a resonant feature when we approach the more strongly interacting superfluid
phase ($V_{\perp }=8\,E_{R}$, and $U\approx J$). The position of this
resonance is well above the frequencies expected for collective oscillations
of the superfluid. We can also rule out excitations to the second Bogoliubov
band since these appear at a much higher frequency.

For a translationally invariant system the appearance of the resonance would
be unexpected for the following reasons. First, the single-phonon excitation
energy is limited by the width of the first Bogoliubov band. For our
parameters this width is $\omega (q_{B})=0.23\,E_{R}/h=0.77$ kHz, where $%
q_{B}$ is the Bragg momentum of the periodic lattice \cite{Kraemer2003}.
This is significantly below the observed resonance close to $1\,E_{R}/h=3.4$
kHz. Secondly, the suppression of the structure factor discussed above for
the weakly interacting Bose gas is based on the f-sum rule \cite{Pines1966}
which is also valid for strongly interacting many-body systems. In liquid
helium for example, inelastic scattering of x-rays and slow neutrons has
shown that long wavelength excitations are suppressed at low temperature
\cite{Gordon1958,Svennson1980}. However, high energy excitations at low
momenta corresponding to multi-phonon states have been observed in liquid
helium \cite{Greytak1970}. Recently it was also suggested that the quantum
depletion of an atomic gas at zero temperature adds a correction to the
dynamic structure factor at low momenta and non-vanishing excitation
energies \cite{Buechler2003}. This result was derived for a three
dimensional system by going beyond the lowest order Bogoliubov expansion. It
thus points out the important role played by the quantum depletion which can
lead to the excitation of two quasi-particles with opposite momenta at a
finite excitation energy.

Recently the parametric excitation of Bogoliubov modes was studied by
solving the time-dependent Gross-Pitaevskii equation for an \ elongated
Bose-Einstein condensate in an optical lattice \cite{Tozzo2004}. In this
simulation the modulation at frequency ${\nu _{mod}}$ drives the parametric
amplification of excitations with momenta $\pm q$ which satisfy the
resonance condition ${\nu _{mod}=2\omega (q)}$. Soon after these two modes
have been amplified, their mean-field interaction with each other and with
the groundstate leads to the population of new modes and a rapid broadening
of the momentum distribution. When ${\nu _{mod}>2\omega (q}_{B})$ the
resonance condition can no longer be satisfied and the excitation strength
decreases. The timescale for the onset of the parametric process decreases
with larger population of the Bogoliubov modes. Therefore the excitation
efficiency should increase with increasing depletion of the condensate, i.e.
for larger $U/J$.
\begin{figure}[tp]
\includegraphics[width=0.8\columnwidth,clip=true]{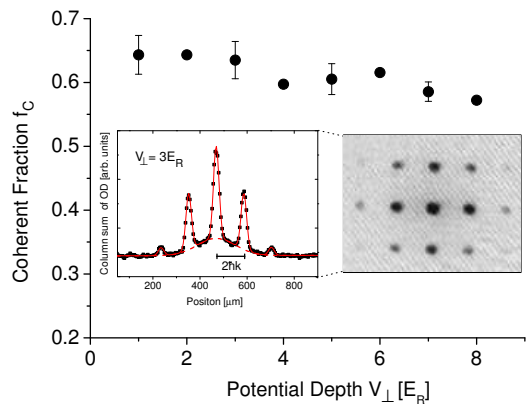}
\caption{Coherent fraction as a function $V_{\perp }$. The inset shows the
absorption image after 10 ms expansion of a condensate initially prepared in
the optical lattice with $V_{\perp }=3\,E_{R}$ (dimensions $570$ $\protect%
\mu $m $\times $ $380$ $\protect\mu $m). The coherent fraction is deduced
from a Gaussian fit (solid plus dashed line) to the column sum of the
optical density (circles), see text. The finite coherent fraction of $65\%$
can be attributed to the quantum depletion in the optical lattice, the
finite temperature of the system and atom-atom scattering during the
expansion from the optical lattice. Error bars are determined by the
statistical error of 4 measurements.}
\label{fig4}
\end{figure}

In order to investigate the role played by the depletion of the condensate
we have measured the coherence properties of the atomic sample both in the
weakly and the more strongly interacting superfluid phase where the
resonance appears. We image the matter wave interference pattern to extract
the coherent fraction \cite{Orzel2001,Greiner2002a,Stoeferle2004}. First we
prepare the Bose gas in the lattice as described above but do not apply our
excitation scheme. Instead, after holding the atoms at the final lattice
depth for $t_{h}=10\,\text{ms}$, we increase all lattice axes rapidly ($%
<40\,\mu \text{s}$) to about $25\,E_{R}$ and then abruptly switch off all
optical and magnetic trapping potentials. This procedure always projects the
state of the superfluid on the same set of Bloch levels. To extract the
number of coherent atoms $N_{coh}$ from the interference pattern, the peaks
at $0\hbar k$, $\pm 2\hbar k$ and $\pm 4\hbar k$ are fitted by Gaussians
(see inset in figure~\ref{fig4}). Incoherent atoms give rise to a broad
gaussian background. Taking this fit as a measure of the number of
incoherent atoms $N_{incoh}$, we calculate the coherent fraction $f_{c}=%
\frac{N_{coh}}{N_{coh}+N_{incoh}}$. In figure~\ref{fig4} the coherent
fraction is shown for increasing values of $V_{\perp }$. The very slow
decrease of $f_{c}$ shows that the system remains coherent and we do not
observe a significant increase in the condensate depletion when preparing
the superfluid in the more strongly interacting regime where $U\approx J$.
Using a Gutzwiller variational calculation we find that, as a result of the
quantum depletion, the condensate fraction is reduced to $0.72$ for the
maximal depth of the optical lattice \cite{Rokhsar1991}.

In conclusion, we have used an optical lattice composed of three orthogonal
standing waves to adjust the level of interactions in a Bose-Einstein
condensed gas and we have investigated the response of this system to the
modulation of the optical lattice amplitude along one axis. In the weakly
interacting regime ($U\ll J$ ) the observed excitation strength remains
close to zero while in the more strongly interacting regime ($U\approx J$) a
broad resonance appears at excitation energies above the width of the first
Bogoliubov band in the optical lattice. Finally we have measured the
depletion of the condensate which remains almost constant for the same
parameters where the resonance appears.

We would like to thank H. P. B\"{u}chler, G. Blatter, F. Dalfovo, M. Kr\"{a}%
mer, L. Pitaevskii, A. S{\o }rensen and C. Tozzo for useful discussions, and
SNF and QSIT for funding.

\end{document}